%% file: main.tex
\pdfoutput = 1
\documentclass[12pt,a4paper]{article}

\usepackage{ifthen}
\newboolean{pdflatex}
\setboolean{pdflatex}{true}

\newboolean{articletitles}
\setboolean{articletitles}{true}

\newboolean{uprightparticles}
\setboolean{uprightparticles}{false}

\usepackage{dcolumn}
\usepackage{placeins}

\newcommand{\longpage}{\enlargethispage{\baselineskip}}

\input{preamble}
\begin{document}

\renewcommand{\thefootnote}{\fnsymbol{footnote}}
\setcounter{footnote}{1}

\input{title-LHCb-PAPER}

\renewcommand{\thefootnote}{\arabic{footnote}}
\setcounter{footnote}{0}

\pagestyle{plain}
\setcounter{page}{1}
\pagenumbering{arabic}

\input{introduction}

\input{selection}

\input{analysis}

\input{systematics2}

\input{conclusions}

\input{acknowledgements}



\addcontentsline{toc}{section}{References}
\bibliographystyle{LHCb}
\bibliography{main}

\end{document}

%% file: preamble.tex
\usepackage{lineno}  
\usepackage{xspace} 

\usepackage{graphicx}  
\usepackage{color}
\usepackage{colortbl}
\graphicspath{{./figs/}} 

\usepackage{amsmath} 
\usepackage{amssymb}
\usepackage{amsfonts}
\usepackage{upgreek} 

\newcommand*\patchAmsMathEnvironmentForLineno[1]{%
\expandafter\let\csname old#1\expandafter\endcsname\csname #1\endcsname
\expandafter\let\csname oldend#1\expandafter\endcsname\csname
end#1\endcsname
 \renewenvironment{#1}%
   {\linenomath\csname old#1\endcsname}%
   {\csname oldend#1\endcsname\endlinenomath}%
}
\newcommand*\patchBothAmsMathEnvironmentsForLineno[1]{%
  \patchAmsMathEnvironmentForLineno{#1}%
  \patchAmsMathEnvironmentForLineno{#1*}%
}
\AtBeginDocument{%
\patchBothAmsMathEnvironmentsForLineno{equation}%
\patchBothAmsMathEnvironmentsForLineno{align}%
\patchBothAmsMathEnvironmentsForLineno{flalign}%
\patchBothAmsMathEnvironmentsForLineno{alignat}%
\patchBothAmsMathEnvironmentsForLineno{gather}%
\patchBothAmsMathEnvironmentsForLineno{multline}%
}

\usepackage{hyperref}    
\usepackage[all]{hypcap} 

\input{lhcb-symbols-def} 

\usepackage{cite} 
\usepackage{mciteplus}

%% file: lhcb-symbols-def.tex
\def\lhcb {LHCb\xspace}
\def\ux85 {UX85\xspace}

\def\babar  {BaBar\xspace}
\def\belle  {Belle\xspace}

\def\cdf    {CDF\xspace}

\def\photos     {\mbox{\textsc{Photos}}\xspace}

\ifthenelse{\boolean{uprightparticles}}%
{

 \def\Pmu         {\ensuremath{\upmu}\xspace}

 \def\Ppi         {\ensuremath{\uppi}\xspace}

 \def\Pphi        {\ensuremath{\upphi}\xspace}

 \def\Ppsi        {\ensuremath{\uppsi}\xspace}

 \def\PDelta      {\ensuremath{\Delta}\xspace}                 
 \def\PXi      {\ensuremath{\Xi}\xspace}                 
 \def\PLambda      {\ensuremath{\Lambda}\xspace}                 
 \def\PSigma      {\ensuremath{\Sigma}\xspace}                 
 \def\POmega      {\ensuremath{\Omega}\xspace}                 
 \def\PUpsilon      {\ensuremath{\Upsilon}\xspace}                 
 
 \def\PB      {\ensuremath{\mathrm{B}}\xspace}                 
                  
 \def\PD      {\ensuremath{\mathrm{D}}\xspace}

 \def\PJ      {\ensuremath{\mathrm{J}}\xspace}                 
 \def\PK      {\ensuremath{\mathrm{K}}\xspace}

 \def\Pb      {\ensuremath{\mathrm{b}}\xspace}                 
 \def\Pc      {\ensuremath{\mathrm{c}}\xspace}

 \def\Pi      {\ensuremath{\mathrm{i}}\xspace}

 \def\Ps      {\ensuremath{\mathrm{s}}\xspace}

}
{

 \def\Pmu         {\ensuremath{\mu}\xspace}

 \def\Ppi         {\ensuremath{\pi}\xspace}

 \def\Pphi        {\ensuremath{\phi}\xspace}

 \def\Ppsi        {\ensuremath{\psi}\xspace}                 
                  
 \mathchardef\PDelta="7101
 \mathchardef\PXi="7104
 \mathchardef\PLambda="7103
 \mathchardef\PSigma="7106
 \mathchardef\POmega="710A
 \mathchardef\PUpsilon="7107
                  
 \def\PB      {\ensuremath{B}\xspace}                 
                  
 \def\PD      {\ensuremath{D}\xspace}

 \def\PJ      {\ensuremath{J}\xspace}                 
 \def\PK      {\ensuremath{K}\xspace}

 \def\Pb      {\ensuremath{b}\xspace}                 
 \def\Pc      {\ensuremath{c}\xspace}

 \def\Pi      {\ensuremath{i}\xspace}

 \def\Ps      {\ensuremath{s}\xspace}

}

\def\mup        {\ensuremath{\Pmu^+}\xspace}
\def\mun        {\ensuremath{\Pmu^-}\xspace} 
\def\mumu       {\ensuremath{\Pmu^+\Pmu^-}\xspace}

\def\squark    {\ensuremath{\Ps}\xspace}

\def\cquark    {\ensuremath{\Pc}\xspace}
\def\cquarkbar {\ensuremath{\overline \cquark}\xspace}

\def\bquark    {\ensuremath{\Pb}\xspace}

\def\pion  {\ensuremath{\Ppi}\xspace}

\def\pip   {\ensuremath{\pion^+}\xspace}
\def\pim   {\ensuremath{\pion^-}\xspace}

\def\kaon  {\ensuremath{\PK}\xspace}
  \def\Kbar  {\kern 0.2em\overline{\kern -0.2em \PK}{}\xspace}

\def\Kz    {\ensuremath{\kaon^0}\xspace}
\def\Kzb   {\ensuremath{\Kbar^0}\xspace}
\def\KzKzb {\ensuremath{\Kz \kern -0.16em \Kzb}\xspace}
\def\Kp    {\ensuremath{\kaon^+}\xspace}
\def\Km    {\ensuremath{\kaon^-}\xspace}

\def\KpKm  {\ensuremath{\Kp \kern -0.16em \Km}\xspace}

\def\Kstarz  {\ensuremath{\kaon^{*0}}\xspace}
\def\Kstarzb {\ensuremath{\Kbar^{*0}}\xspace}
\def\Kstar   {\ensuremath{\kaon^*}\xspace}

  \def\Dbar    {\kern 0.2em\overline{\kern -0.2em \PD}{}\xspace}
\def\D       {\ensuremath{\PD}\xspace}

\def\Dz      {\ensuremath{\D^0}\xspace}
\def\Dzb     {\ensuremath{\Dbar^0}\xspace}
\def\DzDzb   {\ensuremath{\Dz {\kern -0.16em \Dzb}}\xspace}
\def\Dp      {\ensuremath{\D^+}\xspace}
\def\Dm      {\ensuremath{\D^-}\xspace}

\def\DpDm    {\ensuremath{\Dp {\kern -0.16em \Dm}}\xspace}

\def\B       {\ensuremath{\PB}\xspace}
  \def\Bbar    {\kern 0.18em\overline{\kern -0.18em \PB}{}\xspace}

\def\Bz      {\ensuremath{\B^0}\xspace}
\def\Bzb     {\ensuremath{\Bbar^0}\xspace}
\def\Bu      {\ensuremath{\B^+}\xspace}

\def\Bp      {\ensuremath{\Bu}\xspace}

\def\Bd      {\ensuremath{\B^0}\xspace}
\def\Bs      {\ensuremath{\B^0_\squark}\xspace}

\def\Bdb     {\ensuremath{\Bbar^0}\xspace}

\def\jpsi     {\ensuremath{{\PJ\mskip -3mu/\mskip -2mu\Ppsi\mskip 2mu}}\xspace}
\def\psitwos  {\ensuremath{\Ppsi{(2S)}}\xspace}

  \def\Y#1S{\ensuremath{\PUpsilon{(#1S)}}\xspace}

\newcommand{\decay}[2]{\ensuremath{#1\!\to #2}\xspace}         

\def\to                 {\ensuremath{\rightarrow}\xspace}

\def\qsq       {\ensuremath{q^2}\xspace}

\def\CP                {\ensuremath{C\!P}\xspace}

\newcommand{\ACP}{\ensuremath{{\cal A}_{\CP}}\xspace}

\def\BdbToKstmm   {\decay{\Bdb}{\Kstarzb\mup\mun}}

\def\AFB      {\ensuremath{A_{\mathrm{FB}}}\xspace}

\def\AT#1     {\ensuremath{A_{\mathrm{T}}^{#1}}\xspace}           

\def\C#1      {\ensuremath{\mathcal{C}_{#1}}\xspace}                       
\def\Cp#1     {\ensuremath{\mathcal{C}_{#1}^{'}}\xspace}                    
\def\Ceff#1   {\ensuremath{\mathcal{C}_{#1}^{\mathrm{(eff)}}}\xspace}        
\def\Cpeff#1  {\ensuremath{\mathcal{C}_{#1}^{'\mathrm{(eff)}}}\xspace}       
\def\Ope#1    {\ensuremath{\mathcal{O}_{#1}}\xspace}                       
\def\Opep#1   {\ensuremath{\mathcal{O}_{#1}^{'}}\xspace}                    

\newcommand{\tev}{\ensuremath{\mathrm{\,Te\kern -0.1em V}}\xspace}
\newcommand{\gev}{\ensuremath{\mathrm{\,Ge\kern -0.1em V}}\xspace}
\newcommand{\mev}{\ensuremath{\mathrm{\,Me\kern -0.1em V}}\xspace}
\newcommand{\kev}{\ensuremath{\mathrm{\,ke\kern -0.1em V}}\xspace}
\newcommand{\ev}{\ensuremath{\mathrm{\,e\kern -0.1em V}}\xspace}
\newcommand{\gevc}{\ensuremath{{\mathrm{\,Ge\kern -0.1em V\!/}c}}\xspace}
\newcommand{\mevc}{\ensuremath{{\mathrm{\,Me\kern -0.1em V\!/}c}}\xspace}
\newcommand{\gevcc}{\ensuremath{{\mathrm{\,Ge\kern -0.1em V\!/}c^2}}\xspace}
\newcommand{\gevgevcccc}{\ensuremath{{\mathrm{\,Ge\kern -0.1em V^2\!/}c^4}}\xspace}
\newcommand{\mevcc}{\ensuremath{{\mathrm{\,Me\kern -0.1em V\!/}c^2}}\xspace}

\def\mum  {\ensuremath{\,\upmu\rm m}\xspace}

\def\invfb   {\ensuremath{\mbox{\,fb}^{-1}}\xspace}

\newcommand{\chisq}{\ensuremath{\chi^2}\xspace}

\def\gsim{{~\raise.15em\hbox{$>$}\kern-.85em
          \lower.35em\hbox{$\sim$}~}\xspace}
\def\lsim{{~\raise.15em\hbox{$<$}\kern-.85em
          \lower.35em\hbox{$\sim$}~}\xspace}

\def\pt         {\mbox{$p_{\rm T}$}\xspace}

\def\evtgen     {\mbox{\textsc{EvtGen}}\xspace}
\def\pythia     {\mbox{\textsc{Pythia}}\xspace}

\def\geant      {\mbox{\textsc{Geant4}}\xspace}

\def\tell1  {TELL1\xspace}
\def\ukl1   {UKL1\xspace}

\newcommand{\BdToKstarmumu}{\ensuremath{\B^0 \rightarrow K^{*0} \mu^+ \mu^-}\xspace}
\newcommand{\BdToKstarmumubar}{\ensuremath{\Bdb \rightarrow \Kstarzb \mu^+ \mu^-}\xspace}
\newcommand{\BdToJpsiKstar}{\ensuremath{\B^0 \rightarrow J/\psi K^{*0}}\xspace}

\newcommand{\BdTopsitwosKstar}{\ensuremath{\B^0 \rightarrow \psitwos K^{*0}}\xspace}

\newcommand{\BsToJpsiKstar}{\ensuremath{\Bs \rightarrow J/\psi \Kstarzb}\xspace}

\newcommand{\AD}{\ensuremath{\mathcal{A}_\textrm{D}}\xspace}
\newcommand{\AP}{\ensuremath{\mathcal{A}_\textrm{P}}\xspace}
\newcommand{\ARAW}{\ensuremath{\mathcal{A}_\textrm{RAW}}\xspace}

%% file: title-LHCb-PAPER.tex
\begin{titlepage}
\pagenumbering{roman}

\vspace*{-1.5cm}
\centerline{\large EUROPEAN ORGANIZATION FOR NUCLEAR RESEARCH (CERN)}
\vspace*{1.5cm}
\hspace*{-0.5cm}
\begin{tabular*}{\linewidth}{lc@{\extracolsep{\fill}}r}
\ifthenelse{\boolean{pdflatex}}
{\vspace*{-2.7cm}\mbox{\!\!\!\includegraphics[width=.14\textwidth]{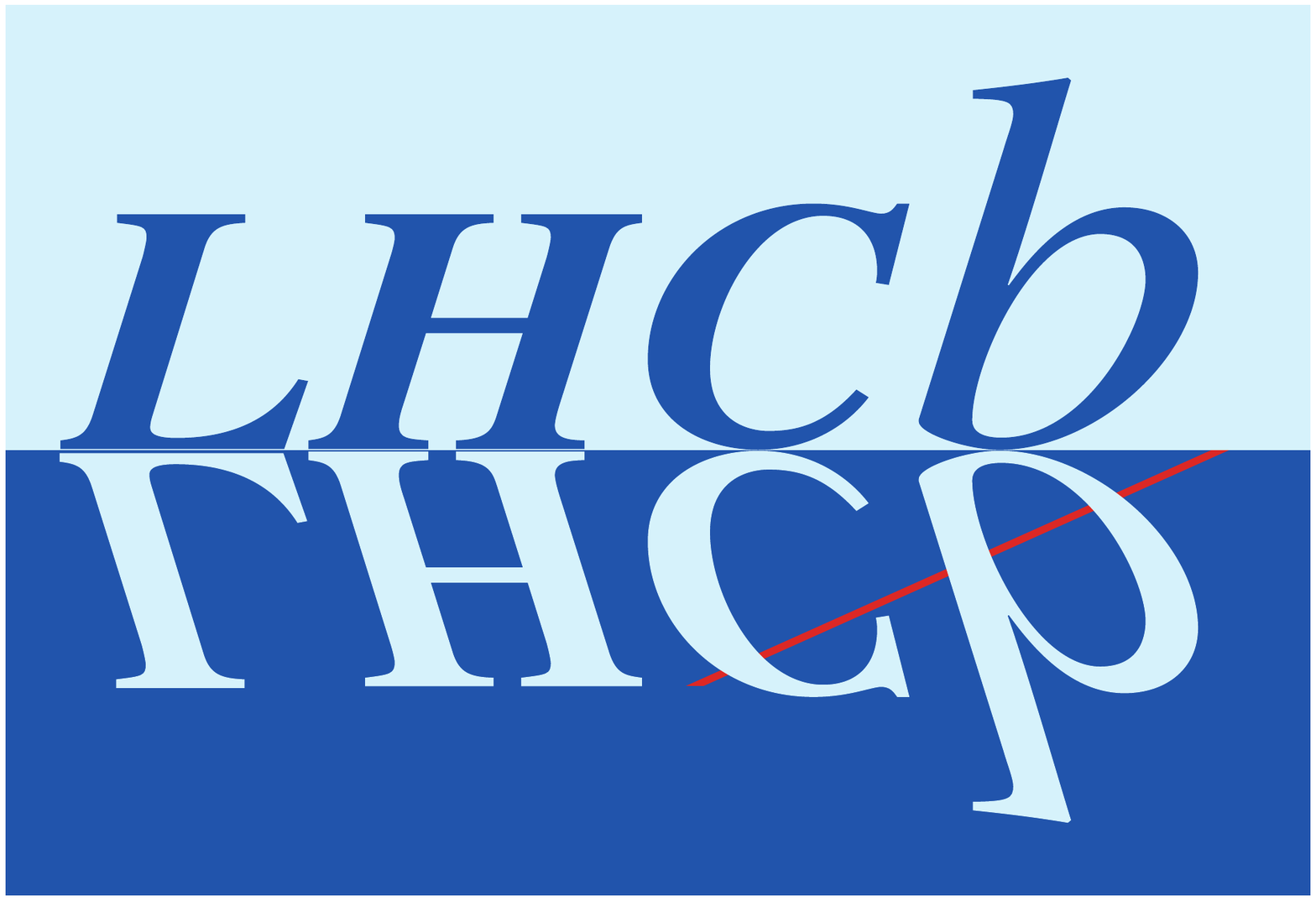}} & &}%
{\vspace*{-1.2cm}\mbox{\!\!\!\includegraphics[width=.12\textwidth]{figs/lhcb-logo.eps}} & &}%
\\
 & & CERN-PH-EP-2012-302 \\ 
 & & LHCb-PAPER-2012-021 \\ 
 & & October 15, 2012 \\
 & & \\
\end{tabular*}

\vspace*{4.0cm}

{\bf\boldmath\huge
\begin{center}
  Measurement of the \CP asymmetry in \BdToKstarmumu decays
\end{center}
}

\vspace*{2.0cm}

\begin{center}
The LHCb collaboration\footnote{Authors are listed on the following pages.}
\end{center}

\vspace{\fill}

\begin{abstract}
  \noindent
A measurement of the \CP asymmetry in \BdToKstarmumu decays is presented, based on 1.0\invfb of $pp$ collision data recorded by the \lhcb experiment during 2011. The measurement is performed in six bins of invariant mass squared of the \mumu pair, excluding the \jpsi and \psitwos resonance regions. Production and detection asymmetries are removed using the \BdToJpsiKstar decay as a control mode. The integrated \CP asymmetry is found to be  $-0.072 \pm 0.040\,(\mbox{stat.}) \pm 0.005\, (\mbox{syst.})$, consistent with the Standard Model.
\end{abstract}

\vspace*{2.0cm}

\begin{center}
  Submitted to Physical Review Letters
\end{center}

\vspace{\fill}

\end{titlepage}

\newpage
\setcounter{page}{2}
\mbox{~}
\newpage

\input{LHCb_authorlist.tex}

\cleardoublepage

%% file: LHCb_authorlist.tex
\centerline{\large\bf LHCb collaboration}
\begin{flushleft}
\small
R.~Aaij$^{38}$, 
C.~Abellan~Beteta$^{33,n}$, 
A.~Adametz$^{11}$, 
B.~Adeva$^{34}$, 
M.~Adinolfi$^{43}$, 
C.~Adrover$^{6}$, 
A.~Affolder$^{49}$, 
Z.~Ajaltouni$^{5}$, 
J.~Albrecht$^{35}$, 
F.~Alessio$^{35}$
% ...

{\footnotesize \it
% ...
$ ^{5}$Clermont Universit\'{e}, Universit\'{e} Blaise Pascal, CNRS/IN2P3, LPC, Clermont-Ferrand, France\\
$ ^{6}$CPPM, Aix-Marseille Universit\'{e}, CNRS/IN2P3, Marseille, France\\
% ...
$ ^{11}$Physikalisches Institut, Ruprecht-Karls-Universit\"{a}t Heidelberg, Heidelberg, Germany\\
% ...
$ ^{33}$Universitat de Barcelona, Barcelona, Spain\\
$ ^{34}$Universidad de Santiago de Compostela, Santiago de Compostela, Spain\\
$ ^{35}$European Organization for Nuclear Research (CERN), Geneva, Switzerland\\
% ...
$ ^{38}$Nikhef National Institute for Subatomic Physics, Amsterdam, The Netherlands\\
% ...
$ ^{43}$H.H. Wills Physics Laboratory, University of Bristol, Bristol, United Kingdom\\
% ...
$ ^{49}$Oliver Lodge Laboratory, University of Liverpool, Liverpool, United Kingdom\\
% ...
$ ^{n}$LIFAELS, La Salle, Universitat Ramon Llull, Barcelona, Spain\\
% ...
}
\end{flushleft}

%% file: introduction.tex
\noindent The decay $B^0 \rightarrow K^{*0} (\rightarrow \Kp \pim) \mu^+ \mu^-$ is a flavour changing neutral current process which proceeds via electroweak loop and box diagrams in the Standard Model (SM) \cite{PhysRevD.61.114028}. The decay is highly suppressed in the SM and therefore physics beyond the SM such as supersymmetry~\cite{Fayet:1976cr} can contribute with a comparable amplitude via gluino or chargino loop diagrams. A number of observables are sensitive to such contributions, including the partial rate of the decay, the \mumu forward-backward asymmetry (\AFB) and the \CP asymmetry (\ACP). The \CP asymmetry for \BdToKstarmumu is defined as
\begin{equation}
 \ACP = \frac{\Gamma(\BdbToKstmm)-\Gamma(\BdToKstarmumu)}{\Gamma(\BdbToKstmm)+\Gamma(\BdToKstarmumu)}
\end{equation}
where $\Gamma$ is the decay rate and the initial flavour of the \B meson is tagged by the charge of the kaon from the \Kstar\ decay. The \CP asymmetry is predicted to be of the order $10^{-3}$ in the SM~\cite{Bobeth:2008ij,Altmannshofer:2008dz}, but is sensitive to physics beyond the SM that changes the operator basis by modifying the mixture of the vector and axial-vector components~\cite{Ali:1998sf,Bobeth:2011nj}. Some models that include new phenomena enhance the observed \CP asymmetry up to $\pm 0.15$~\cite{Alok:2011gv}. The theoretical prediction within a given model has a small error as the form factor uncertainties, which are the dominant theoretical errors for the decay rate, cancel in the ratio.

The \CP asymmetry in \BdToKstarmumu decays has previously been measured by the Belle~\cite{Wei:2009zv} and BaBar~\cite{Babar:2012vwa} collaborations, with both results consistent with the SM.  The \lhcb collaboration has recently demonstrated its potential in this area with the most precise measurement of \AFB~\cite{Aaij:2011aa}, and in this Letter, the measurement of the \CP asymmetry by \lhcb is presented.

%% file: selection.tex
The \lhcb detector~\cite{Alves:2008zz} is a single-arm forward
spectrometer covering the pseudorapidity range $2<\eta <5$, designed
for the study of particles containing \bquark or \cquark quarks. The
detector includes a high precision tracking system consisting of a
silicon-strip vertex detector surrounding the $pp$ interaction region,
a large-area silicon-strip detector located upstream of a dipole
magnet with a bending power of about $4{\rm\,Tm}$, and three stations
of silicon-strip detectors and straw drift-tubes placed
downstream. The combined tracking system has a momentum resolution
$\Delta p/p$ that varies from 0.4\% at 5\gevc to 0.6\% at 100\gevc,
and an impact parameter (IP) resolution of 20\mum for tracks with high
transverse momentum (\pt). Charged hadrons are identified using two
ring-imaging Cherenkov detectors. Photon, electron and hadron
candidates are identified by a calorimeter system consisting of
scintillating-pad and pre-shower detectors, an electromagnetic
calorimeter and a hadronic calorimeter. Muons are identified by a 
system composed of alternating layers of iron and multiwire
proportional chambers. The trigger consists of a hardware stage, based
on information from the calorimeter and muon systems, followed by a
software stage which makes use of a full event reconstruction.

The simulated events used in this analysis are produced using the \pythia 6.4 generator \cite{Sjostrand:2006za}, with a choice of parameters specifically configured for \lhcb \cite{LHCb-PROC-2011-005}. The \evtgen package \cite{Lange:2001uf} describes the decay of the particles and the \geant toolkit \cite{Allison:2006ve, *Agostinelli:2002hh} simulates the detector response, implemented as described in Ref.~\cite{LHCb-PROC-2011-006}. QED radiative corrections are generated with the \photos package \cite{Golonka:2005pn}.

The events used in the analysis are selected by a dedicated muon hardware trigger and then by one or more of a set of different muon and topological software triggers~\cite{LHCb-PUB-2011-017,LHCb-PUB-2011-016}. The hardware trigger requires the muons have \pt greater than 1.48\gevc, and the software trigger requires that one of the final state particles to have both $\pt > 0.8\gevc$ and IP with respect to all $pp$ interaction vertices $> 100\mum$~\cite{LHCb-PUB-2011-016}. Triggered candidates are subject to the same two-stage selection as that used in Ref.~\cite{Aaij:2011aa}. The first stage is a cut-based selection, which includes requirements on the \Bz candidate's vertex fit \chisq, flight distance and invariant mass, and each track's impact parameters with respect to any interaction vertex, \pt and polar angle.
Background from misidentified kaon and pion tracks is removed using information from the particle identification (PID) system, and muon tracks are required to have hits in the muon system. Finally, the production vertex of the \Bz candidate must lie within 5 mm of the beam axis in the transverse directions, and within 200 mm of the average interaction position in the beam ($z$) direction.

In the second stage, the candidates must pass a multivariate selection that uses a boosted decision tree (BDT)~\cite{Breiman} that implements the AdaBoost algorithm~\cite{AdaBoost}. This is a tighter selection which takes into account other variables including the decay time and flight direction of the \Bz candidates, the \pt of the hadrons, measures of the track and vertex quality, and PID information for the daughter tracks. For the rest of the Letter, the inclusion of charge conjugate modes is implied unless explicitly stated.

In order to obtain a clean sample of \BdToKstarmumu decays, the $c\bar{c}$ resonant decays \BdToJpsiKstar and \BdTopsitwosKstar are removed by excluding events with \mumu invariant mass, $m_{\mumu}$, satisfying $2.95 < m_{\mumu} < 3.18\gevcc$ or $3.59 < m_{\mumu} < 3.77\gevcc$. If $m_{\Kp\pim\mumu} < 5.23\gevcc$, then the vetoes are extended downwards by $0.15\gevcc$ to remove the radiative tails of the resonances. Backgrounds involving misidentified particles are vetoed using cuts on the masses of the \Bz and \Kstarz mesons and the $\mup \mun$ pair, as well as using the PID information for the daughter particles. These include $\Bs \rightarrow \Pphi\mumu$ candidates in which a kaon has been misidentified as a pion, \BdToJpsiKstar candidates where a hadron is swapped with a muon and $\Bp \rightarrow \Kp \mumu$ candidates which combine with a random low momentum pion. The vetoes are described fully in Ref.~\cite{Aaij:2011aa}. \ACP may be diluted by \BdToKstarmumu candidates with the kaon and pion misidentified as each other, which is estimated as 0.8\% of the \BdToKstarmumu yield using simulated events.
All \Bz candidates must have a mass in the range $5.15-5.80\gevcc$, the tight low mass edge of this window serves to remove background from partially reconstructed $B$ meson decays. All \Kstarz candidates must have an invariant mass of the kaon-pion pair within $0.1\gevcc$ of the nominal $\kaon^{*0}$($892$) mass. A proton veto, using PID information from a neural network, is also applied to remove background from $\Lambda_{b}$ decays, where a proton in the final state is misidentified as a kaon or pion in the \BdToKstarmumu decay.

Approximately 2\% of selected events contain two \BdToKstarmumu candidates which have tracks in common. The majority of these candidates arise from swapping the assignment of the kaon and pion hypothesis. As the charges of the kaon and pion tag the flavour of the $B$ meson these duplicate candidates can bias the measured value of \ACP. This is accounted for by randomly removing one of the two candidates from the sample. This process is repeated many times over the full sample with a different random seed in each case and the average measured value of \ACP is taken as the result.

%% file: analysis.tex
An accurate measurement of \ACP requires that the differences in the production rates~($R$) of \Bd/\Bdb mesons and detection efficiencies~($\epsilon$) between the \BdToKstarmumu and \BdToKstarmumubar modes be accounted for. Assuming all asymmetries are small the raw measured asymmetry may be expressed as
\begin{equation}
 \ARAW = \ACP + \kappa\AP + \AD,
\end{equation}
where the production asymmetry, which is of the order of 1\%~\cite{Aaij:2012qe}, is defined as ${\AP \equiv \left[ R \left( \Bzb \right) - R\left( \Bz \right) \right]/\left[ R \left( \Bzb \right) + R\left( \Bz \right) \right]}$ and the detection asymmetry is ${\AD \equiv \left[ \epsilon \left( \Bzb \right) - \epsilon\left( \Bz \right) \right]/\left[ \epsilon \left( \Bzb \right) + \epsilon\left( \Bz \right) \right]}$. The production asymmetry is diluted through \Bd-\Bdb oscillations by a factor $\kappa$
\begin{equation}
\kappa \equiv \frac{\int_0^\infty \epsilon(t) e^{-\Gamma t} \cos \Delta m t \, \mathrm{ d}t}{\int_0^\infty \epsilon(t) e^{-\Gamma t} \,\mathrm{ d}t},
\end{equation}
where $t$, $\Gamma$, and $\Delta m$ are the decay time, mean decay rate, and mass difference between the light and heavy eigenstates of the \Bz meson respectively. The quantity \AD is dominated by the $\Kp\pim/\Km\pip$ detection asymmetry which arises due to left-right asymmetries in the \lhcb detector and different interactions of positively and negatively charged tracks with the detector material. The left-right asymmetry is cancelled by taking an average with equal weights of the \CP asymmetries measured in two independent data samples with opposite polarities of the \lhcb dipole magnet. These data samples correspond to 61\% and 39\% of the total data sample.

 The production and interaction asymmetries are corrected for using the \BdToJpsiKstar decay mode as a control channel. Since \BdToKstarmumu and \BdToJpsiKstar decays have the same final state and similar kinematics, the measured raw asymmetry for ${\B \rightarrow \jpsi K^{*0}}$ decays may be simply expressed as $\ARAW\left(\BdToJpsiKstar\right) = \kappa \AP + \AD$, in the absence of a \CP asymmetry. \BdToJpsiKstar proceeds via a $b \to c\cquarkbar s$ transition, as does the decay mode $\Bp \to \jpsi \Kp$, and hence should have a \CP asymmetry similar to $\ACP(\Bp \to \jpsi \Kp) = (1 \pm 7) \times 10^{-3}$~\cite{PDG2012,Abazov:2008gs}. For this analysis, it is assumed that $\ACP(\BdToJpsiKstar) = 0$. The \CP asymmetry in \BdToKstarmumu decays is then calculated as
\begingroup
\small
\begin{equation}
\ACP = \ARAW\left(\BdToKstarmumu\right) - \ARAW\left(\BdToJpsiKstar\right).
\end{equation}
\endgroup
Non-cancelling asymmetries due to differences between the kinematics of \BdToKstarmumu and \BdToJpsiKstar decays are considered as systematic effects.

The full data sample, containing approximately 900 \BdToKstarmumu signal decays, is split into the six bins of \mumu invariant mass squared (\qsq) used by the \lhcb, \belle, and \cdf angular analyses \cite{Aaij:2011aa, Wei:2009zv, Aaltonen:2011ja}. An additional bin of $1<\qsq<6\gevgevcccc$ is used, to be compared to the theoretical prediction in Ref.~\cite{Altmannshofer:2008dz}. The \BdToJpsiKstar data sample contains approximately 104000 signal decays with $3.04<\qsq<3.16\gevgevcccc$. The values of \ACP are measured using a simultaneous unbinned maximum-likelihood fit to the \BdToKstarmumu and \BdToJpsiKstar invariant mass distributions in the range $5.15-5.80\gevcc$. The simultaneous fit in each \qsq bin spans eight data samples, split between the initial particles \Bz and \Bzb, the decay modes \BdToKstarmumu and \BdToJpsiKstar, and magnet polarity, where the \BdToJpsiKstar sample is common to all \qsq bins. This fit returns two values of \ACP, one for each magnet polarity, and an average with equal weights is made to find the value of \ACP in that \qsq bin. An integrated value of \ACP over all \qsq is also calculated.

\begin{figure}[htb]
 \centering
\includegraphics[width=\textwidth]{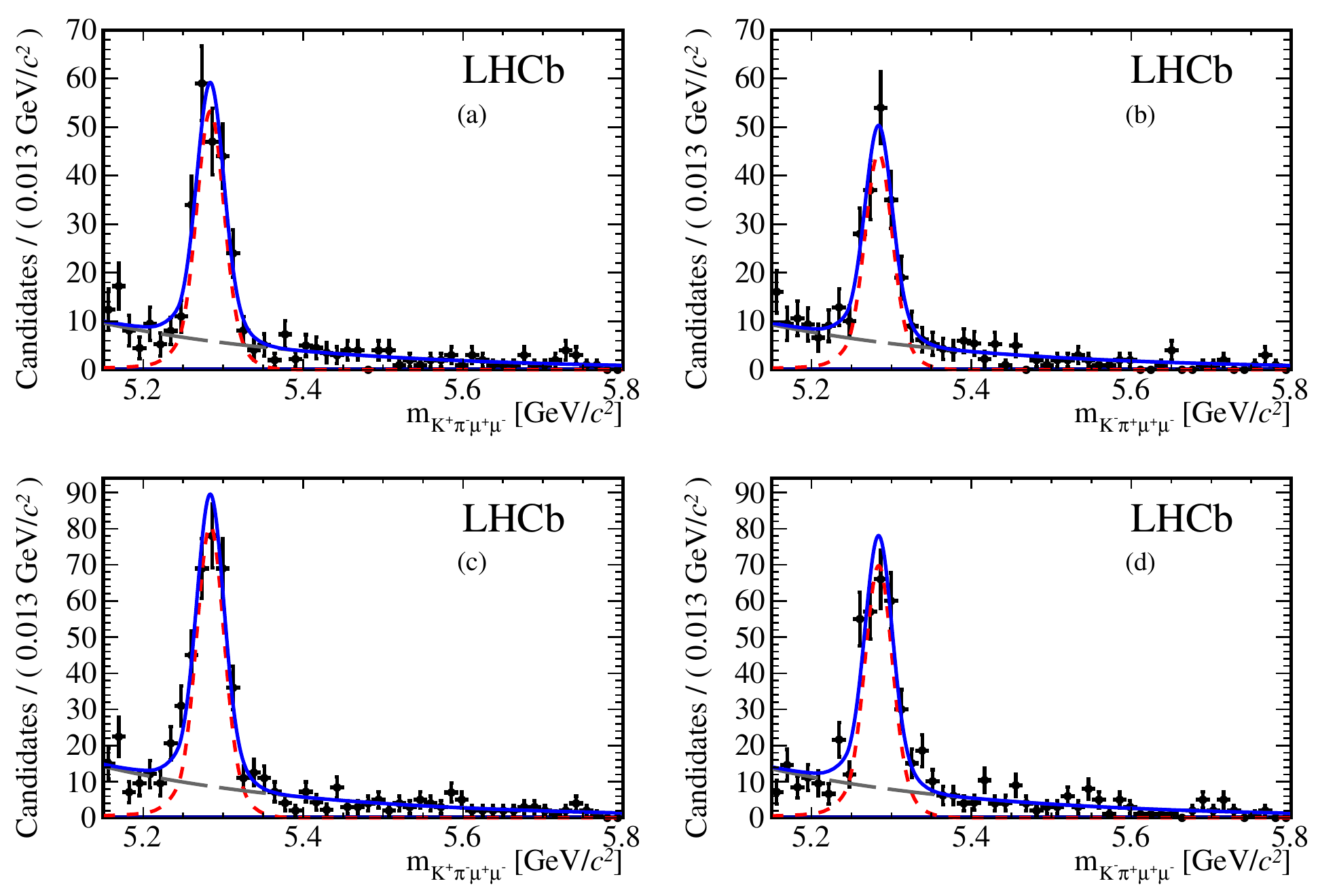}
\caption{Mass fits for \BdToKstarmumu decays used to extract the integrated \CP asymmetry. The curves displayed are the full mass fit (blue, solid), the signal peak (red, short-dashed), and the background (grey, long-dashed). The mass fits on the top row correspond to the (a) \Bz and (b) \Bzb decays for one magnet polarity, while the bottom row shows the mass fits for (c) \Bz and (d) \Bzb for the reverse polarity.}
\label{fig:BdToKstarmumu_Integrated_Fit}
\end{figure}

The signal invariant mass distributions for the \BdToKstarmumu and \BdToJpsiKstar decays are modelled using the sum of two Crystal Ball functions~\cite{Skwarnicki:1986xj} with common peak and tail parameters but different widths. The values of the tail parameters are determined from fits to simulated events and fixed in the fit. Combinatorial background arising from the random misassociation of tracks to form a \Bz candidate is modelled using an exponential function. The \BdToJpsiKstar fit also accounts for a peaking \BsToJpsiKstar contribution, which has the same shape as the signal and an expected yield which is $(0.7 \pm 0.2) \%$ of that of \BdToJpsiKstar~\cite{Aaij:2012nh}. In the simultaneous fit, the signal shape is the same for the two modes, but the signal and background yields and the exponential background parameter may vary. Figure~\ref{fig:BdToKstarmumu_Integrated_Fit}
shows the mass fit to the \BdToKstarmumu decay in the full \qsq range.

%% file: systematics2.tex
Many sources of systematic uncertainty cancel in the difference of the raw asymmetries between \BdToKstarmumu and \BdToJpsiKstar decays and in the average of \CP asymmetries measured using data recorded with opposite magnet polarities. However, systematic uncertainties can arise from residual non-cancelling asymmetries due to the different kinematic behaviour of \BdToKstarmumu and \BdToJpsiKstar decays. The effect is estimated by reweighting \BdToJpsiKstar candidates so that their kinematic variables are distributed in the same way as for \BdToKstarmumu candidates. The value of $\ARAW\left(\BdToJpsiKstar\right)$ is then calculated for these reweighted events and the difference from the default value is taken as the systematic uncertainty. This procedure is carried out separately for a number of quantities including the $p$, \pt, and pseudorapidity of the \Bd and the \Kstarz mesons. The total systematic uncertainty associated to the different kinematic behaviour of the two decays is calculated by adding each individual contribution in quadrature. This is
conservative, as many of the variables are correlated.

The random removal of multiple candidates discussed above also introduces a systematic uncertainty on \ACP. The uncertainty on the mean value of \ACP from the ten different random removals is taken as the systematic uncertainty.

The forward-backward asymmetry in \BdToKstarmumu decays~\cite{Aaij:2011aa}, which varies as a function of \qsq, causes positive and negative muons to have different momentum distributions. Different detection efficiencies for positive and negative muons introduce an asymmetry that cannot be accounted for by the \BdToJpsiKstar decay, which does not have a comparable forward-backward asymmetry. The selection efficiencies for positive and negative muons are evaluated using muons from \jpsi decay in data and the resulting asymmetry in the selected \BdToKstarmumu sample is calculated in each \qsq bin.

A number of possible effects due to the choice of model for the mass fit are considered. The signal model is replaced with a sum of two Gaussian distributions and a possible difference in the mass resolution for \BdToKstarmumu and \BdToJpsiKstar decays is investigated by allowing the width of the \BdToKstarmumu signal peak to vary in a range of 0.7$-$1.3 times that of the \BdToJpsiKstar model. As a further cross-check, \ACP is calculated using a weighted average of the measurements from the six \qsq bins and the result is found to be consistent with that obtained from the integrated dataset. 
\longpage

\begin{table}[tb]
\centering
\caption{Systematic uncertainties on \ACP, from residual kinematic asymmetries, muon asymmetry, choice of signal model, and the modelling of the mass resolution, for each \qsq bin. The total uncertainty is calculated by adding the individual uncertainties in quadrature.}
\begin{tabular}{|r@{\qsq}l|D{.}{.}{1.4}|D{.}{.}{1.4}|D{.}{.}{1.4}|D{.}{.}{1.4}|D{.}{.}{1.4}|D{.}{.}{1.4}|}  \hline
                        \multicolumn{2}{|c|}{}& \multicolumn{5}{c|}{Sources of systematic uncertainties} &   \\ \hline
\multicolumn{2}{|c|}{} & \multicolumn{1}{c|}{multiple}& \multicolumn{1}{c|}{residual} & \multicolumn{1}{c|}{$\mu^{\pm}$ detection} & \multicolumn{1}{c|}{signal} & \multicolumn{1}{c|}{mass} & \multicolumn{1}{c|}{}\\
\multicolumn{2}{|c|}{\qsq~region ($\gevgevcccc$)} & \multicolumn{1}{c|}{cands.}& \multicolumn{1}{c|}{asymmetries} & \multicolumn{1}{c|}{asymmetry} & \multicolumn{1}{c|}{model} & \multicolumn{1}{c|}{resol.} & \multicolumn{1}{c|}{Total} \\ \hline 
$0.05<~$&$<2.00$	& 0.002 & 0.007 & 0.005 & 0.005 & 0.001 & 0.010 \\
$2.00<~$&$<4.30$	  	& 0.006 & 0.007 & 0.006 & 0.007 & 0.010 & 0.016 \\ 
$4.30<~$&$<8.68$	  	& 0.004 & 0.003 & 0.006 & 0.004 & 0.003 & 0.010 \\ 
$10.09<~$&$<12.86$ 	& 0.003 & 0.007 & 0.009 & 0.001 & 0.002 & 0.011 \\ 
$14.18<~$&$<16.00$	  	& 0.001 & 0.006 & 0.007 & 0.001 & 0.001 & 0.009 \\ 
$16.00<~$&$<20.00$	  	& 0.003 & 0.005 & 0.003 & 0.003 & 0.009 & 0.012 \\ \hline 
$1.00<~$&$<6.00$	  	& 0.001 & 0.006 & 0.005 & 0.002 & 0.003 & 0.009 \\ \hline 
$0.05<~$&$<20.00$	  	& 0.002 & 0.002 & 0.005 & 0.001 & 0.001 & 0.005 \\ \hline 
\end{tabular}
\label{tab:systematicUncertainties}
\end{table}

%% file: conclusions.tex
\begin{figure}[tb]
 \centering
\includegraphics[width=0.7\textwidth]{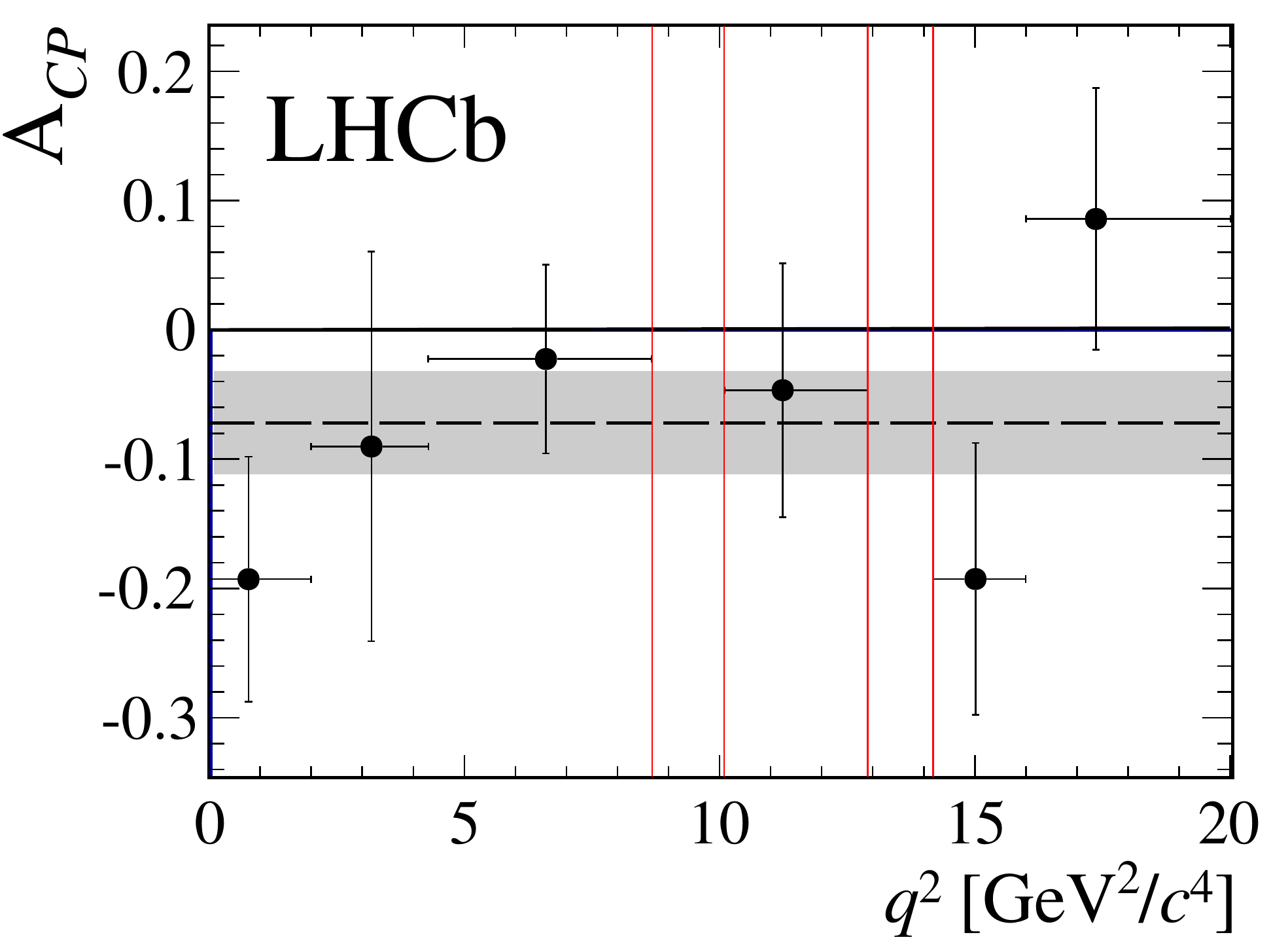}
\caption{Fitted value of \ACP in \BdToKstarmumu decays in bins of the \mumu invariant mass squared (\qsq). The red vertical lines mark the charmonium vetoes. The points are plotted at the mean value of \qsq in each bin. The uncertainties on each \ACP value are the statistical and systematic uncertainties added in quadrature. The dashed line corresponds to the \qsq integrated value, and the grey band is the 1$\sigma$ uncertainty on this value.}
\label{fig:ACPvsqsq}
\end{figure}

The results of the full \ACP fit are presented in Table~\ref{tab:results} and Figure~\ref{fig:ACPvsqsq}. The raw asymmetry in \BdToJpsiKstar decays is measured as
\begin{equation*}
 \ARAW \left( \BdToJpsiKstar \right) = -0.0110 \pm 0.0032 \pm 0.0006.
\end{equation*}
where the first uncertainty is statistical and the second is systematic. The \CP asymmetry integrated over the full \qsq  range is calculated and found to be
\begin{equation*}
 \ACP \left( \BdToKstarmumu \right) = -0.072 \pm 0.040 \pm 0.005.
\end{equation*}
The result is consistent with previous measurements made by \belle~\cite{Wei:2009zv}, \ACP$(B \rightarrow \Kstar l^+ l^-) = -0.10 \pm 0.10 \pm 0.01$, and \babar~\cite{Babar:2012vwa}, \ACP$(B \rightarrow \Kstar l^+ l^-) = 0.03 \pm 0.13 \pm 0.01$.  This measurement is significantly more precise than all other measurements of \ACP in \BdToKstarmumu decays to date.

\begin{table}[!tb]
\centering
\caption{Values of \ACP for \BdToKstarmumu in the \qsq bins used in the analysis.}
\begin{tabular}{|r@{\qsq}l|r@{$\pm$}l|D{.}{.}{1.4}|D{.}{.}{1.4}|D{.}{.}{1.4}|D{.}{.}{1.4}|}
	\hline 
\multicolumn{2}{|c|}{}  &  \multicolumn{2}{c|}{signal}  &   & \multicolumn{1}{c|}{statistical} & \multicolumn{1}{c|}{systematic}  & \multicolumn{1}{c|}{total}  \\ 
  \multicolumn{2}{|l|}{\qsq~region ($\gevgevcccc$)}  & \multicolumn{2}{c|}{yield} & \multicolumn{1}{c|}{$~~~\ACP~~~$} & \multicolumn{1}{c|}{uncertainty} & \multicolumn{1}{c|}{uncertainty} & \multicolumn{1}{c|}{uncertainty} \\ \hline 
$0.05<~$&$<2.00$	 &$168$&$15$ & -0.196 & 0.094 & 0.010  & 0.095 \\
$2.00<~$&$<4.30$	 &$72$&$11$ & -0.098 & 0.153 & 0.016  & 0.154 \\
$4.30<~$&$<8.68$	 &$266$&$19$ & -0.021 & 0.073 & 0.010  & 0.075 \\
$10.09<~$&$<12.86$	 &$157$&$15$ & -0.054 & 0.097 & 0.011 & 0.098 \\
$14.18<~$&$<16.00$	 &$116$&$12$ & -0.201 & 0.104  & 0.009 & 0.104 \\
$16.00<~$&$<20.00$	 &$128$&$13$ & 0.089 & 0.100 &  0.012 & 0.101 \\ \hline
$1.00<~$&$<6.00$	 &$194$&$17$ & -0.058 & 0.064 &  0.009 & 0.064 \\ \hline
$0.05<~$&$<20.00$	 &$904$&$35$ & -0.072 & 0.040 & 0.005 & 0.040 \\ \hline

\end{tabular}
\label{tab:results}
\end{table}

\clearpage

%% file: acknowledgements.tex
\section*{Acknowledgements}

\noindent We express our gratitude to our colleagues in the CERN
accelerator departments for the excellent performance of the LHC. We
thank the technical and administrative staff at the LHCb
institutes. We acknowledge support from CERN and from the national
agencies: CAPES, CNPq, FAPERJ and FINEP (Brazil); NSFC (China);
CNRS/IN2P3 and Region Auvergne (France); BMBF, DFG, HGF and MPG
(Germany); SFI (Ireland); INFN (Italy); FOM and NWO (The Netherlands);
SCSR (Poland); ANCS/IFA (Romania); MinES, Rosatom, RFBR and NRC
``Kurchatov Institute'' (Russia); MinECo, XuntaGal and GENCAT (Spain);
SNSF and SER (Switzerland); NAS Ukraine (Ukraine); STFC (United
Kingdom); NSF (USA). We also acknowledge the support received from the
ERC under FP7. The Tier1 computing centres are supported by IN2P3
(France), KIT and BMBF (Germany), INFN (Italy), NWO and SURF (The
Netherlands), CIEMAT, IFAE and UAB (Spain), GridPP (United
Kingdom). We are thankful for the computing resources put at our
disposal by Yandex LLC (Russia), as well as to the communities behind
the multiple open source software packages that we depend on.